\documentclass[]{pasj01}

\begin{document} 
\Received{}
\Accepted{}

\title{Observations of V404 Cygni during the 2015 outburst by the Nasu telescope array at 1.4\,GHz}

\author{Kuniyuki \textsc{Asuma}\altaffilmark{1,2}}
\altaffiltext{1}{Faculty of Science and Engineering, Waseda University, 3-4-1 Okubo, Shinjuku-ku, Tokyo 169-8555}
\altaffiltext{2}{Asaka High School, Saiwaicho 3-13-65, Asaka-shi, Saitama 351-0015}

\author{Kotaro \textsc{Niinuma}\altaffilmark{3,}$^*$}
\altaffiltext{3}{Graduate School of Sciences and Technology for Innovation, Yamaguchi University, Yoshida 1677-1, Yamaguchi 753-8512}
\email{niinuma@yamaguchi-u.ac.jp}

\author{Kazuhiro \textsc{Takefuji}\altaffilmark{4,5}}
\altaffiltext{4}{Kashima Space Research Center, National Institute of Information and Communications Technology, 893-1 Hirai, Kashima-shi, Ibaraki 314-8501}
\altaffiltext{5}{Japan Aerospace Exploration Agency, 3-1-1 Yoshinodai, Chuo-ku, Sagamihara, Kanagawa 229-8510, Japan}

\author{Takahiro \textsc{Aoki}\altaffilmark{5}}
\altaffiltext{6}{The Research Institute for Time Studies, Yamaguchi University, Yoshida 1677-1, Yamaguchi 753-8511}

\author{Sumiko \textsc{Kida}\altaffilmark{1}}

\author{Hirochika \textsc{Nakajima}\altaffilmark{1}}

\author{Kimio \textsc{Tsubono}\altaffilmark{1}}

\author{Tsuneaki \textsc{Daishido}\altaffilmark{1,7}}
\altaffiltext{7}{School of Education, Waseda University, 1-104 Totsukamachi, Shinjuku-ku, Tokyo 169-8050}


\maketitle

\begin{abstract}

Waseda University Nasu telescope array is a spatial fast Fourier transform (FFT) interferometer consisting of eight linearly aligned antennas with 20\,m spherical dishes. This type of interferometer was developed to survey transient radio sources with an angular resolution as high as that of a 160\,m dish with a field of view as wide as that of a 20\,m dish. We have been performing drift-scan-mode observations, in which the telescope scans the sky around a selected declination as the earth rotates. 
The black hole X-ray binary V404 Cygni underwent a new outburst in 2015 June after a quiescent period of 26 years. Because of the interest in black hole binaries, a considerable amount of data on this outburst at all wavelengths was accumulated. 
Using the above telescope, we had been monitoring V404 Cygni daily from one month before the X-ray outburst, and two radio flares at 1.4\,GHz were detected on June 21.73 and June 26.71. The flux density and time-scale of each flare were $313\pm30$ mJy and $1.50\pm0.49$ days, $364\pm30$ mJy and $1.70\pm0.16$\,days, respectively. 
%
We have also confirmed the extreme variation of radio spectra within a short period by collecting other radio data observed with several radio telescopes. Such spectral behaviors are considered to reflect the change in the opacity of the ejected blobs associated with these extreme activities in radio and X-ray.  Our 1.4\,GHz radio data are expected to be helpful for studying the physics of the accretion and ejection phenomena around black holes.

\end{abstract}

\section{Introduction}\label{sec1}
The Nasu telescope array operates as a spatial FFT interferometer. In this scheme, an antenna array and an FFT processor produce a direct image of an arriving radio source. A number of topics dealt with here have been reviewed by Thompson et al. (2017). The idea of a spatial FFT interferometer was proposed in 1984 and demonstrated using an eight-element test telescope at the campus of Waseda University \citep{Daishido 1984, Daishido 1987}. Afterwards, two $8\times 8$ pilot two-dimensional\, (2D) arrays were constructed; the larger one (overall size of $20\,\mathrm{m}\times 20\,\mathrm{m}$) was designed for surveying transient radio sources, and the smaller one ($1.2\,\mathrm{m}\times 1.2\,\mathrm{m}$) was designed for the observation of cosmic microwave background fluctuations \citep{Asuma 1991,Daishido 1991}. The larger array, consisting of $8\times8$ dishes of 2.4\, m diameter, was used to study the direct imaging scheme by observing bright radio sources. By executing a 2D spatial FFT of the signals from each antenna element, we were able to obtain $8\times8$ pixel images of the sky every 50\, ns \citep{Nakajima 1993, Otobe 1994}. We have also developed a large-scale interferometric array consisting of eight linearly aligned antennas with 20\, m spherical dishes at the Jiyu-Gakuen Nasu Farm \citep{Daishido 2000}. Using this Nasu telescope array, we have been searching for transient radio sources, such as an unknown strong flare lasting for a few days \citep{Niinuma 2007}.

V404 Cygni is known for the detection of exhibit extremely bright and variable activity, which was detected by the X-ray satellite Ginga in 1989 \citep{Makino(1989)}. It was first reported as the bright X-ray nova GS 2023+338; however, subsequent investigations showed that V404 Cygni was in an outburst state, and consequently is the optical counterpart of an X-ray source (Marsden 1989). V404 Cygni is considered to be a binary system containing a stellar-mass black hole that is the closest to us among all such known systems.  The binary system has a $\sim 10\, M_\odot$ black hole and a $\sim 1\, \MO$ companion with an orbital period of 6.5 days. Later radio parallax measurements determined its distance to be $\sim 2.4\,\mathrm{kpc}$ \citep{Miller-Jones(2009)}.

On 2015 June 15, V404 Cygni went into outburst again. Gamma-ray bursts from V404 Cygni were first detected by Swift/BAT \citep{Barthelmy 2015} and also reported by MAXI \citep{Negoro 2015}. These  early  alerts  triggered  follow-up observations at all wavelengths. Independently of this, using Waseda University Nasu telescope array, we had been monitoring V404 Cygni daily from one month before the outburst. We were therefore able to observe a sudden increase in activity in V404 Cygni at 1.4\, GHz. Immediately after the observations, we reported preliminary results in two short notes \citep{Tsubono(2015a), Tsubono(2015b)}.

Here, first we describe the Nasu telescope array, which was constructed on the basis of the novel idea of using a spatial FFT interferometer, and then we report our observations of the V404 Cygni outburst. Finally, we discuss the physical implications of our data, considered along with other observational results. 

\section{Spatial FFT interferometer}\label{sec2}

Suppose that incident radio waves arrive at antennas placed at equally spaced grid points in a plane. By spatially Fourier transforming the electric field sampled by each antenna, we can obtain a map of the incident field in $k$-space. Thus, the combination of an antenna array and a real-time spatial FFT processor acts as a digital lens, which produces a direct image of  the incoming waves. We call this type of detector a spatial FFT interferometer. This type of interferometer is suitable for surveying transient radio sources.

\subsection{Principle of spatial FFT interferometer}

Assume that an infinite number of antennas are distributed continuously on an $x-y$ plane, and then a plane radio wave arrives at the antenna plane. The electric field on the antenna located at $\boldsymbol{r}$ is written in the form
    \begin{eqnarray}
    \boldsymbol{E}(\boldsymbol{r},t)=\boldsymbol{E_0}e^{i(\boldsymbol{k_0r}-\omega_0 t)},
    \end{eqnarray}

where $\boldsymbol{E_0}$, $\boldsymbol{k_0}$ and $\omega_0$ are the amplitude of the electric field, the wave-number vector and the angular frequency of the incident wave, respectively. Note that $\boldsymbol{E}$, $\boldsymbol{E_0}$, $\boldsymbol{k_0}$ and $\boldsymbol{r}$ are all two-dimensional vectors on the $x-y$ plane. If we execute the Fourier transformation of $\boldsymbol{E}(\boldsymbol{r},t)$ with respect to the spatial coordinate $\boldsymbol{r}$, we obtain
 
    \begin{eqnarray}
    \widetilde{ \boldsymbol{E}(\boldsymbol{r},t)}
    &=&\left(\frac{1}{2\pi}\right)^2\int^{\infty}_{-\infty}\int^{\infty}_{-\infty}\boldsymbol{E_0}e^{i(\boldsymbol{k_0r}-\omega_0 t)}e^{-i\boldsymbol{kr}}d\boldsymbol{r}
    \label{eq:fourie} \nonumber\\
    &=&\boldsymbol{E_0}e^{-i\omega_0 t}\delta(\boldsymbol{k}-\boldsymbol{k_0}).
    \end{eqnarray}

\begin{figure}[tb]
        \begin{center}
        \includegraphics[width=0.95\linewidth]{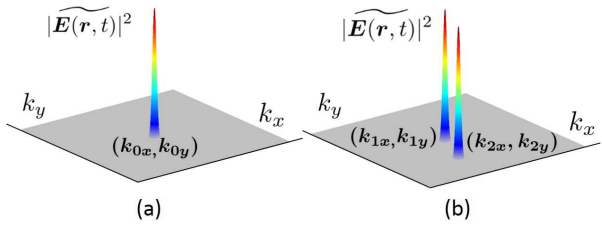}
        \end{center} 
                \caption{(a) Image on the $\boldsymbol{k}$-plane formed by a point source in the   $\boldsymbol{k_0}$-direction. (b) Images formed by two independent point sources.}
        \label{fig:point source}
     \end{figure}

Therefore, by finding the peak of $\widetilde{\boldsymbol{E}}$ in $\boldsymbol{k}$-space, we can determine the direction $\boldsymbol{k_0}$ of the incident wave. Figure \ref{fig:point source} illustrates the peaks appearing in $\boldsymbol{k}$-space originating from the incident waves from (a) a point source and (b) two independent sources. The incident angle $\theta$ with respect to the vertical axis is calculated from the following relation
    \begin{eqnarray}
    \sin \theta=\frac{\lambda_0}{2\pi}k_0,
        \label{eq:sin theta}
    \end{eqnarray}
where $\lambda_0$ is the wavelength of the incident wave. 

%
If the distribution of antennas is finite, the $\delta$-function in equation\,(\ref{eq:fourie}) is replaced by a broadened function. For simplicity, let the antennas be distributed inside the area $|x|\le b/2$ and $|y|\le b/2$; then equation\,(\ref{eq:fourie}) with a definite integral interval yields a spatial frequency response function with bandwidth
    \begin{eqnarray}
    \Delta k\sim \frac{2\pi}{b}.
    \end{eqnarray}
From equation (\ref{eq:sin theta}), we can derive the angular resolution as
    \begin{eqnarray}
    \Delta\theta \sim \frac{\lambda_0}{b}. 
    \end{eqnarray}
More practically, if the antennas are distributed discretely, not continuously, on the grid points with equal intervals $d$, the sampling theorem implies that the output of equation\,(\ref{eq:fourie}) has a periodic pattern with interval
    \begin{eqnarray}
    \Delta k=\frac{2\pi}{d}.
    \end{eqnarray}
Thus, if the incident angle $\theta$ satisfies $|\theta|\ll 1$, the field of view is given by
    \begin{eqnarray}
    \Delta\theta \sim \frac{\lambda_0}{d}.
    \end{eqnarray}

As described above, the angular resolution is given by the overall size of the antenna array, while the field of view is determined by the spacing between the antenna elements.  We can thus design the configuration of the spatial FFT interferometer appropriately according to the purpose of the observation.

Most of the traditional Fourier synthesis telescopes in use today were designed to obtain fine images of radio sources using  a relatively small number of antennas by choosing minimum- redundancy baselines or an arbitrary configuration of antennas. They are indirect imaging systems that use correlators and integrators.  On the other hand,  the spatial FFT interferometer,  in which each antenna element is fixed in the maximum redundancy position, generates real-time images of radio sources at the Nyquist sampling rate. Moreover,  an $N$-element spatial FFT interferometer requires $N\log_2N$ multipliers,  while an $N$-element Fourier synthesis interferometer requires $N(N-1)$ correlators. This means that the spatial FFT type is more economical than the correlator type when the number of elements $N$ is large \citep{Daishido 1984}.

\subsection{Nasu telescope array}
    \begin{figure}[htb]
    \begin{center}
    \includegraphics[width=0.85\linewidth]{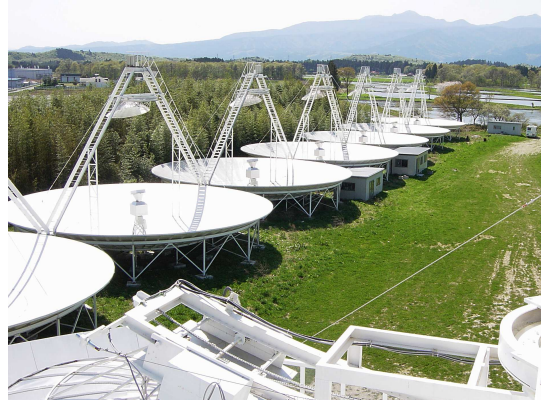}
    \end{center}
     \caption{Nasu telescope array with 20\,m spherical antennas at the Jiyu-Gakuen Nasu Farm.}
    \label{fig:nasu_array}
    \end{figure}

\begin{figure}[htb]
    \begin{center}
    \includegraphics[width=0.55\linewidth]{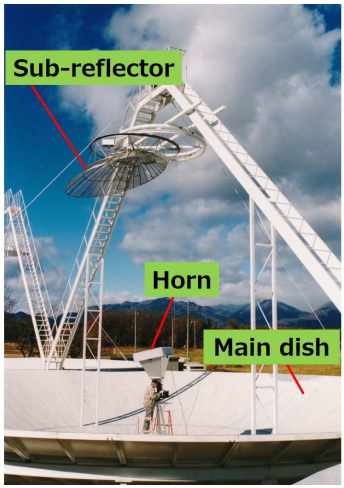}
    \end{center}
    \caption{Principal parts of each antenna: main dish, sub-reflector, and feed horn.}
    \label{fig:nasu_structure}
    \end{figure}

%
The Nasu telescope array is located at the Jiyu-Gakuen Nasu Farm in Tochigi Prefecture, 160 km north of Tokyo, at latitude ${36}^\circ55'41.3''$ north and longitude ${139}^\circ58'54.3''$ east. The telescope is a one-dimensional array of eight antennas with a spacing of 21 m, aligned from east to west (E-W). The main part of each antenna is a spherical dish with a diameter of 20\, m. A photograph of the 20 m antennas is shown in figure \ref{fig:nasu_array}. The radius of curvature and the aperture diameter are both 20\, m; that is, each antenna is a ${60}^\circ$ spherical cap. For an incoming radio wave, the spherical surface of the main reflector and a Gregorian sub-reflector form a focal point at the input of the feed horn\ (see figure \ref{fig:nasu_structure}). The asymmetrical 3D surface of the sub-reflector was specially designed to compensate for the aberrations caused by the spherical reflector \citep{Daishido 2000}.

The main dish is fixed on the ground, whereas the sub-reflector and feed horn can be moved mechanically. The sub-reflector is located ${5}^\circ$ from the vertex axis of the spherical reflector, and thus, the elevation angle is fixed to ${85}^\circ$. By rotating the sub-reflector synchronously with the feed horn in azimuth, the antenna covers the sky area in a declination zone of ${32}^\circ \le \delta \le {42}^\circ$, which is 7.0\% of the entire sky.

    \begin{figure}[htb]
    \begin{center}
    \includegraphics[width=1.0\linewidth]{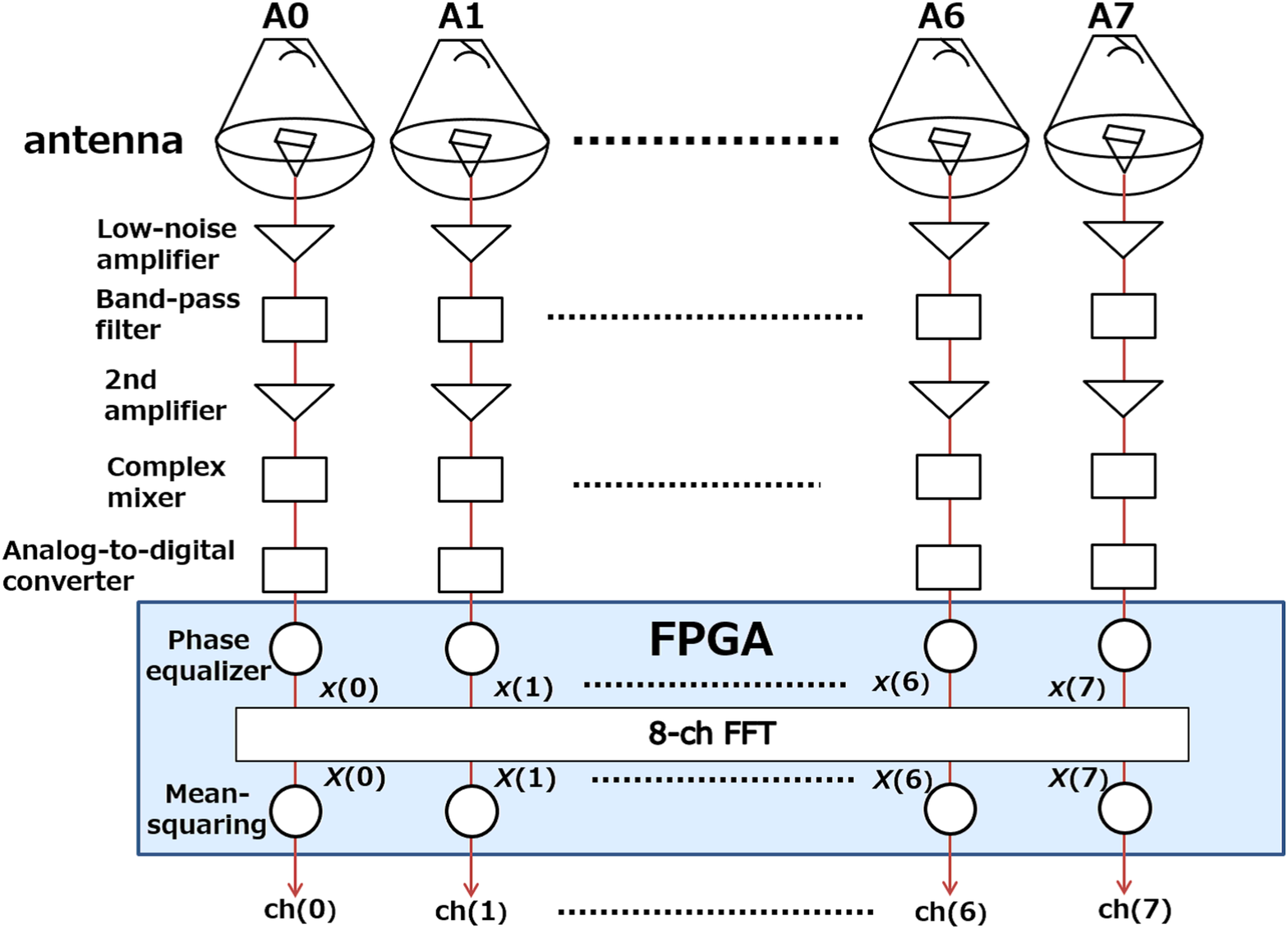}
    \end{center}
    \caption{Block diagram of the Nasu telescope array including the data processing system.}
    \label{fig:diagram}
    \end{figure}

A block diagram of the Nasu telescope array, including the data processing system, is shown in figure \ref{fig:diagram}. The detector system operates at a center frequency of 1.415\, GHz with a bandwidth of 20\, MHz. The front-end signal from each feed horn is fed into an HEMT low-noise amplifier with a noise temperature of $\sim$40\,K. After being filtered within a 1.405--1.425\, GHz bandwidth and amplified by 40\, dB, the signal is downconverted to 20\, MHz by a complex mixer with two 1.415\, GHz references in quadrature. The complex signal is sampled by an 8-bit analog-to-digital converter every 50\, ns. 
The digitized signal is subsequently processed by an FPGA-based processor. The phase fluctuation arising from the temperature-dependent delay in the transmission line is compensated for by the phase equalizer. The 8-ch complex signal $x(n)\  (n=0,1,...,7)$ is then spatially Fourier transformed to obtain $X(k)$ as follows

\begin{eqnarray}
X(k)={\displaystyle \sum_{n=0}^{7}x(n)\exp(-i2\pi\frac{kn}{8})}.
\end{eqnarray}
Finally, the time-averaged vector amplitude
\begin{eqnarray}
\mathrm{ch}(k)=\overline{|X(k)|^2} \label{eq:ch(k)}
\end{eqnarray}
is calculated and sent to the storage device.  

The main parameters of the Nasu telescope array are summarized in table \ref{table:parameters}.
\begin{table}[htbp]
 \caption{Main parameters of the Nasu telescope array.}
     \label{table:parameters}
 \begin{center}
  \begin{tabular}{lll}
    \hline
Location      & Latitude ${36}^\circ55'41.3''$ north   &       \\
  &Longitude ${139}^\circ58'54.3''$ east\\
Number of antennas & Eight with 21\,m spacing (E-W)& \\
Main reflector      &Fixed 20\,m spherical  dish  &       \\
Sub-reflector      &Asymmetrical Gregorian type    &       \\ 
Angular resolution      &${0.1}^\circ$ (E-W)   &       \\ 
Field of view (HPBW)      &${0.8}^\circ$    &       \\ 
Declination coverage      &${32}^\circ \le \delta \le {42}^\circ$    &       \\
Frequency range      &1.415 $\pm$ 0.01\,GHz    &       \\
Nyquist frequency      &20\,MHz    &       \\
    \hline
  \end{tabular}
 \end{center}
\end{table}

\subsection{Data analysis}

The directivity of the spatial FFT interferometer is determined by the configuration of the antenna array. On the basis of the characteristics of the directivity pattern, two analysis methods have been developed: one is direct imaging and the other is correlation analysis.
\subsubsection{Directivity patterns}
In a spatial FFT interferometer, each ch($k$) in equation\,(\ref{eq:ch(k)}) has its own directivity pattern, which we will derive here. First, neglecting the curvature of the antenna dish, we can obtain  the directivity $F_C$ of a single round dish for an incoming electric wave $e^{i\boldsymbol{k_0r}}$ as follows
    \begin{eqnarray}
    F_C&=&\int^{}_{r\le a}e^{i\boldsymbol{k_0r}}dS \nonumber\\
       &=&\pi a^2\frac{2J_1(k_0a)}{k_0a},
    \end{eqnarray}
where a is the radius of the dish. Next, the array factor $F_A$ for $N$-element linearly aligned antennas with an equal spacing of $\boldsymbol{d}$ is calculated as
    \begin{eqnarray}
   F_A(k)&=&\sum_{n=0}^{N-1}e^{-i\frac{2\pi}{N}nk}e^{in\boldsymbol{k_0d}}\nonumber\\
    &=&e^{\frac{i}{2}(N-1)(\boldsymbol{k_0d}-\frac{2\pi}{N}k)
    }\frac{\sin (\kappa_0-k)\pi}{\sin(\kappa_0-k) \frac{\pi}{N}},
    \end{eqnarray}
where we defined the modified incident direction $\kappa_0$ as $\kappa_0=\frac{N\boldsymbol{k_0d}}{2\pi}$.  
    \begin{figure}[htb]
    \begin{center}
    \includegraphics[width=0.8\linewidth]{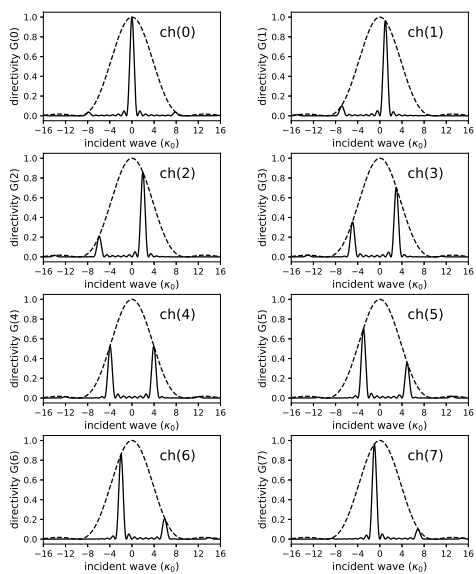}
    \end{center}
    \caption{Normalized directivity pattern $G(k)$ for each ch($k$) $(k=0, 1, 2, ...,7)$ as a function of $\kappa_0$.  Here we assume that $\boldsymbol{k_0}$ is parallel to $\boldsymbol{d}$ and $d=2a$.  The solid line shows the antenna pattern for  each ch($k$) and the broken line shows that of a single dish.}
    \label{fig:pattern}
    \end{figure}
\renewcommand{\figurename}{Fig.}
\noindent
Then we can obtain the following normalized directivity pattern $G(k)$ for each ch($k$) $(k=0, 1, 2, ..., N-1)$: 
    \begin{eqnarray}
    G(k)&=&|F_CF_A(k)|^2 \nonumber\\
    &=&\biggl|\frac{2J_1(k_0a)}{k_0a}\frac{\sin (\kappa_0-k)\pi}{\sin(\kappa_0-k) \frac{\pi}{N}}\biggr|^2.
    \end{eqnarray}
Assuming that $N=8$, $\boldsymbol{k_0}$ is parallel to $\boldsymbol{d}$ and $d=2a$, the antenna pattern $G(k)$ as a function of $\kappa_0$ is shown in figure\ \ref{fig:pattern}.

\subsubsection{Direct imaging}
\begin{figure}[htb]
    \begin{center}
    \includegraphics[width=1.05\linewidth]{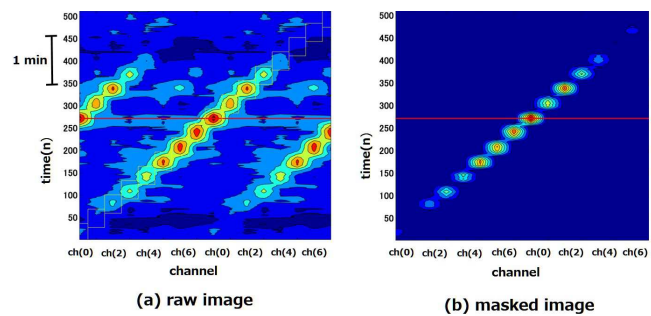}
    \end{center}
    \caption{Example of direct image produced by the bright radio source 4C 33.57 (1.1\,Jy at 1.4\,GHz).   The integration time of ch($k$) is 0.6\,s. (a) Raw image appearing in the same two sets of ch($k$). (b)  Masked view of the same image.}
    \label{fig:direct_image}
    \end{figure}

A direct image can be obtained by making a contour map on the array block of ch($k$) in equation\,(\ref{eq:ch(k)}).  Figure \ref{fig:direct_image} shows an example of a direct image obtained from the bright radio source 4C 33.57 (1.1\,Jy at 1.4\,GHz; \cite{Condon(1998)}).  As shown in figure \ref{fig:direct_image} (a), the same two sets of ch($k$) reduce the complexity of the folded pattern arising from the Fourier transformation.  Furthermore, we can obtain the clearer image, as shown in figure \ref{fig:direct_image} (b) by masking the unnecessary part of the array before making the contour map.

%
In principle, the direct imaging of a radio source shows intensity variations with a short duration. For this purpose, however, an accurate calibration of the image is crucial, although the optimum method that should be used is still under investigation. Nevertheless, we can apply direct imaging to distinguish artificial noise, such as radio-frequency interference (RFI), from genuine astronomical radio signals. Direct imaging cannot be applied to very faint radio sources.
\subsubsection{Correlation analysis}
As shown in figure \ref{fig:pattern}, each ch($k$) has its own antenna pattern. We can therefore find signals buried in an output stream by using correlation analysis or a pattern-matching method. This technique is a very powerful tool in the data analysis of gravitational waves, especially for the chirp waves generated by a binary coalescence \citep{Abbott(2016)}. In correlation analysis, the observed signal-to-noise ratio (SNR) for each ch($k$) is formulated as follows (see, for example, \cite{Creighton(2011)}):
    \begin{figure}[htb]
    \begin{center}
    \includegraphics[width=1.0\linewidth]{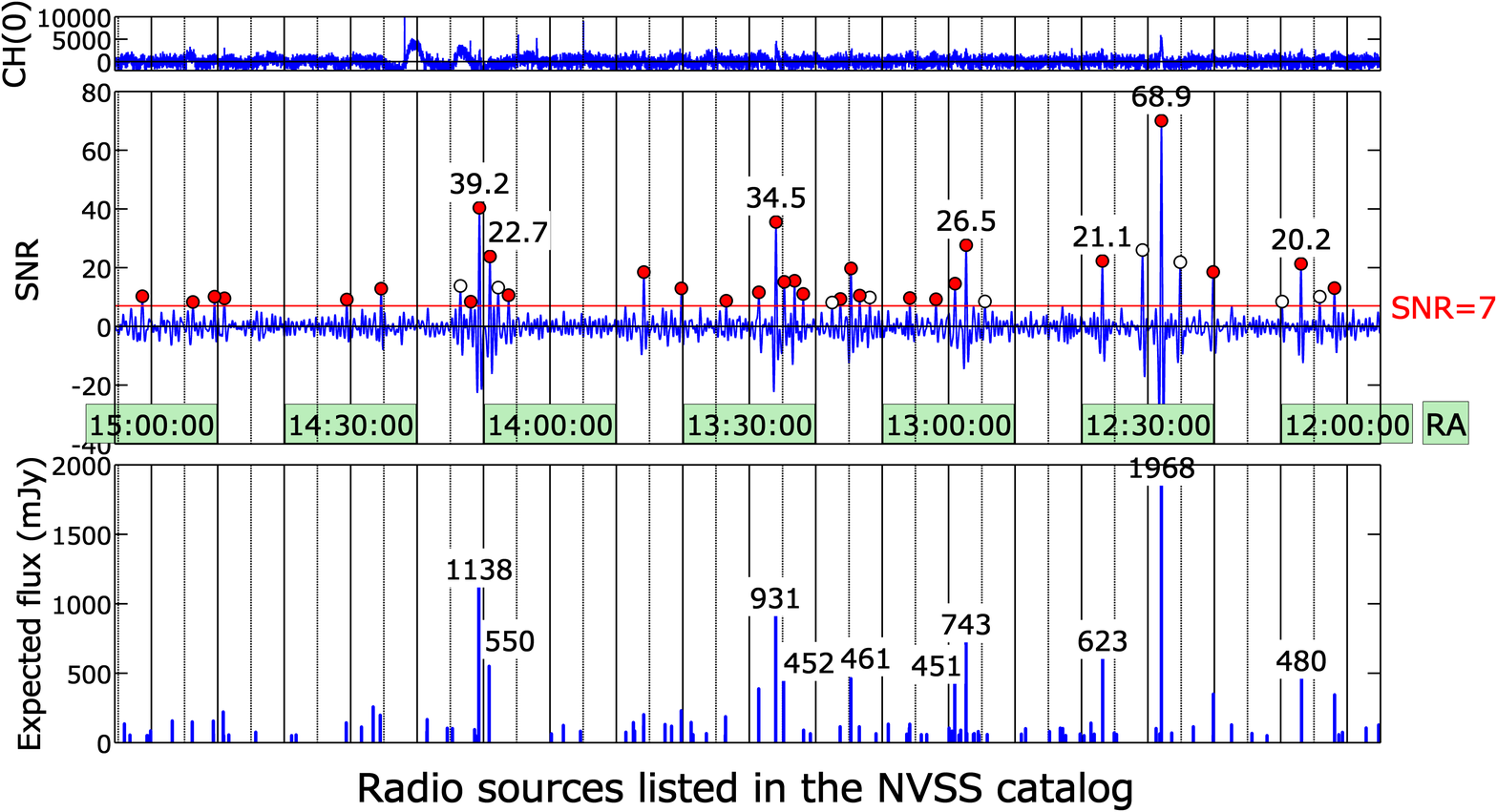}
    \end{center}
    \caption{
    Example of three-hour data analyzed by the pattern-matching method. The upper figure shows the raw data of ch(0). The middle figure shows the combined SNR of the 8-ch output data. SNR peaks larger than 7 are marked with a small circle; the figures by the peaks are observed SNRs. Peaks with an open circle show the side-lobe from strong sources. The lower figure shows the expected flux density of the radio sources listed in the NRAO VLA Sky Survey (NVSS) catalog \citep{Condon(1998)} taking account of the difference in the declinations between the source and the antenna position.}
    \label{fig:pattern_matching}
    \end{figure}
\begin{eqnarray}
\mathrm{SNR}=4\mathrm{Re}\int_{0}^{+\infty} \frac{\tilde{x}(f)\hat{s}^*(f)e^{2\pi ift}}{P_{\mathrm{one}}(f)}df.
\end{eqnarray}
In our case, $\tilde{x}(f)$ is the Fourier transform of ch($k$) with respect to time $t$, $\hat{s}(f)$ is the normalized Fourier transform  of the antenna pattern with respect to time $t$, and $P_{\mathrm{one}}(f)$ is the one-sided power spectral density of ch($k$). This expression is essentially a noise-weighted correlation of the anticipated wave form with the actual data.

An example of three-hour data analyzed by the pattern- matching method is shown in figure \ref{fig:pattern_matching}. The upper figure shows the raw data of ch(0). The middle figure shows the reduced  SNR of the 8-ch output data. SNR peaks larger than 7 are marked with a small circle, and the values shown at the peaks are the observed SNRs. The lower figure shows the expected flux density of the radio sources listed in the NVSS catalog, taking account of the difference in the declinations between the source and the antenna positions. We obtained good agreement between the expected and observed source distributions and their intensities.

The observed one-sigma noise level was $\sim 20$\,mJy when the averaging time of ch($k$) is 0.6\,s. Thus, this method is effective even for rather weak radio sources.

\section{Observations and results of V404 Cygni}
The Nasu telescope array started an observation run on 2015 May 18, monitoring the sky region inside the declination zone of $33.8^\circ \pm 0.4^\circ$ in the drift-scan mode, in which the telescope scans the sky around a selected declination as the earth rotates. V404 Cygni is located at \timeform{20h 24m 03.83s} in right ascension (RA) and ${33.9}^\circ$ in declination (Dec); thus, this source was within our survey area. 
    \begin{figure}[htb]
    \begin{center}
    \includegraphics[width=1.00\linewidth]{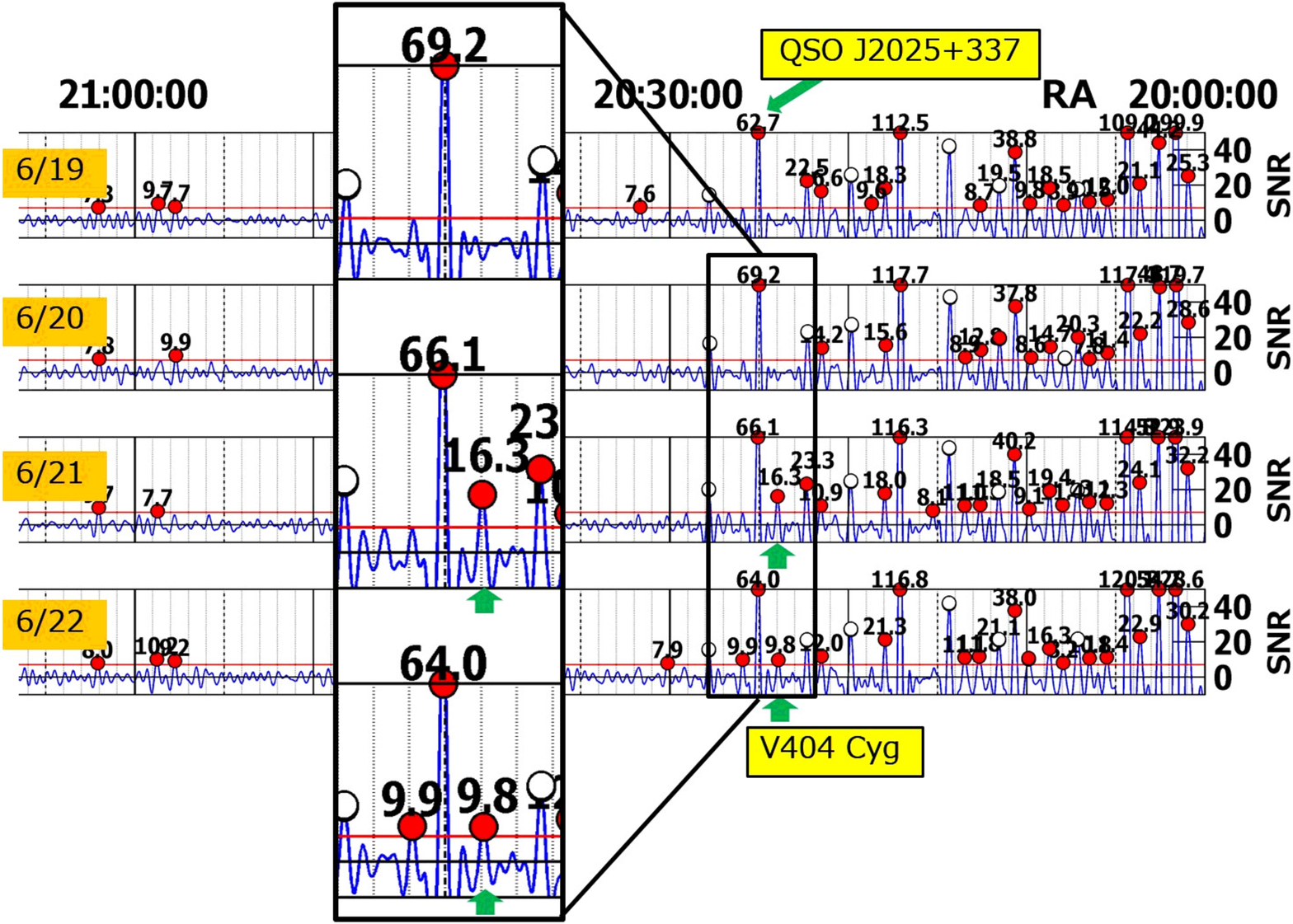}
    \end{center}
    \caption{
    Daily records of the one-hour SNR data from June 19 to June 22. An enlarged view of a part of the figure is also shown.  Peaks with an open circle show the side-lobe from strong sources.  On June 21 17:24 UT, we found a significant peak at a position near V404 Cygni, although on June 20 17:28 UT we could not observe any noticeable signal in the vicinity. A nearby radio source, QSO J2025+337, was chosen as a calibration star, whose flux density is 1268\, mJy at 1.4\, GHz \citep{Condon(1998)}. }
    \label{fig:quick_data}
    \end{figure}

\begin{figure}[htb]
    \begin{center}
    \includegraphics[width=1.00\linewidth]{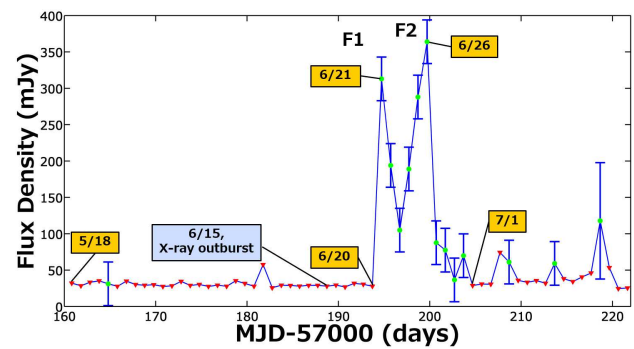}
    \end{center}
    \caption{
    Daily variation of the detected flux density around V404 Cygni during the two months from May 18 (MJD 57160) to July 18 UT (MJD 57221) including the period of the V404 Cygni outburst. Except for the 10 days from June 21 (MJD 57194) to June 30 UT (MJD 57203), the detected signal was below the detection limit (filled red triangle). These detection limits can sometimes be high, mainly due to bad weather at the observation site.  The date of the X-ray outburst is also shown here.}
    \label{fig:long_light_curve}
    \end{figure}

Figure \ref{fig:quick_data} shows the daily records of the one-hour SNR data calculated by the correlation analysis method. In the plot, the horizontal axis shows the RA of the target sky region. Four days of data from June 19 to June 22 are plotted here to demonstrate how to find substantial changes in the brightness among the daily records. On 2015 June 21, 17:24 UT,  we found a significant peak at a position near V404 Cygni, although on June 20 17:28 UT, we could not observe any noticeable signal in the vicinity. A nearby radio source, QSO J2025+337, whose flux density is 1268 mJy at 1.4\, GHz, was chosen  as a calibrator \citep{Condon(1998)}. The calibrated flux density of the detected signal on June 21 was 313 $\pm$ 30 mJy \citep{Tsubono(2015a)}. Since the center position of V404 Cygni and the calibrator is almost on the observation line, the difference in directivity was not considered here. The daily variation of the detected flux density during the two months from May 18 (MJD 57160) to July 18 UT (MJD 57221), which includes the period of the V404 Cygni outburst, is plotted  in figure \ref{fig:long_light_curve}. Except for the 10 days from June 21 (MJD 57194) to June 30 UT (MJD 57203), the detected signal was below the detection limit. 

On June 21 17:24 UT, we found a significant peak of 313 $\pm$ 30 mJy at the position of V404 Cygni. Then, the observed flux density decreased to 105 $\pm$ 30 mJy on June 23 17:17 UT. However, the flux density recovered from that point and increased to 364 $\pm$ 30 mJy on June 26 17:05 UT. On June 27 17:01 UT, we observed fast decay of the intensity \citep{Tsubono(2015b)}. After this rapid decay, the flux density seemed to slowly decay. These characteristics appeared to be similar to those of the radio decay curve reported for the 1989 outburst of V404 Cygni \citep{[Han(1992)}.

\section{Discussions}
As mentioned in Section \ref{sec1}, \citet{Barthelmy 2015} and \citet{Negoro 2015} reported that Swift-BAT and MAXI-GSC captured the huge X-ray outburst, which preceded the radio outburst, in the black-hole X-ray binary object V404 Cygni. After this X-ray observation and in response to the alert about its detection, follow-up observations were performed by ground-based telescopes, including not only radio telescopes but also other telescopes such as optical telescopes. On the other hand, fortunately, we had been monitoring the region at a declination line of $33.8^\mathrm{\circ}\pm0.4^\mathrm{\circ}$, at which V404 Cygni is located, every day since 2015 May 18, which was almost one month before the X-ray outburst and during which time V404 Cygni was in the quiescent state in the 1.4\, GHz radio band.

Although an accurate understanding of the flux increase in the flaring phase of non-thermal synchrotron emission is essential for investigating the coupling between the accretion state and the jet ejection mechanism, which is one of the most important questions, it is quite difficult to obtain early-stage information of the flare especially for radio observation, not only at higher frequencies but also at lower frequencies. Because of their smaller field of view (FoV), most radio telescopes have not performed unbiased surveys, but have carried out follow-up observations of transient phenomena triggered by X-ray/$\gamma$-ray observations by all-sky monitoring satellites.

As shown in figure \ref{fig:long_light_curve}, it was revealed that the radio behavior of this source was clearly very quiescent around June 15 when the first X-ray outburst was recognized in this object.
It is easy to specify the date on which the radio outburst occurred at 1.4\, GHz with a time resolution of one day. This daily light curve clearly shows two radio outbursts at 2015 June 21.73 (the first flare: F1) and 26.71 (the second flare: F2). 
%
%
In order to estimate the time scale of flares, we performed the structure function analysis (e.g., \cite{1985ApJ...296...46S}). The first-order structure function $SF(\tau)$ is described as follows: 
\begin{eqnarray}
	SF(\tau)=\Big<[S(t)-S(t+\tau)]^2\Big>
\end{eqnarray}
where $S$ is the observed flux density, and $\tau$ is the time lag. 
$SF(\tau)$ derived from daily light curve at 1.4GHz gives us a characteristic time scale of $\sim3$ days, which is thought to be corresponding to the approximate duration of each individual flare i.e., F1 and F2. 
For more quantitative estimation of the time scale of F1 and F2, we fitted the function of double exponential form as follows to the time profile of single flare \citep{2010ApJ...722..520A}:
\begin{eqnarray}
	F(t) = F_c\Big(e^\frac{t_0-t}{t_r} + e^\frac{t-t_0}{t_d}\Big) + F_b \label{eq:e14}
\end{eqnarray} 
where, $F_b$ is a baseline level of light curve in quiescent state, $F_c$ is an amplitude coefficient, $t_0$ is estimated time of the flare peak, $t_r$ and $t_d$ are e-folding times in rise and decay phase, respectively. 
Here, we define the duration of individual flare as $T_{flare}(\equiv t_r+t_d$), and $T_{flare}$ of F1 and F2 derived from the well-fitted results with Eq.(\ref{eq:e14}) are $1.50\pm0.49$ days and $1.70\pm0.16$ days, respectively.
Surprisingly, the flux density of F1 drastically increased by more than 10-times within a day compared to that in the quiescent state (e.g., in 2015 June 20.73). On the other hand, it is difficult to understand the situation in which $t_d$ of F2 is clearly shorter than $t_r$ in the framework of the synchrotron regime. 
Therefore, the flux peak at 1.4\, GHz in F2 possibly appeared between June 25.71 and 26.71.
%
%

$T_{flare}$ suggests that the size of the emitting region where both radio outbursts occurred is estimated to be $R<\sim3.9\times10^{15}\mathrm{cm}$, which is derived from the light-crossing time ($R<c\Delta t$) under the assumption of non-relativistic phenomena. 
%
Other radio follow-up observations with high time resolution but short duration clearly revealed intraday variability from this object (e.g., \cite{Gandhi(2017),Tetarenko(2017),Tetarenko(2019)}) especially at higher radio frequencies ($> 10$GHz). \citet{Tetarenko(2017)} \& \citet{Tetarenko(2019)} have reported that the radio fluxes at higher frequencies showed high variability for a few hours in both outbursts (i.e., F1 \& F2).  However, the variabilities seen at lower frequencies of less than a few GHz seem to be much slower compared to those seen at higher frequencies. 
Based on the magnetic field strength $B$ discussed in \citet{Chandra(2017)}, the time scale of synchrotron cooling at 1.4\, GHz should be
\begin{equation}
t_\mathrm{syn}~\approx~3.5 \left(\frac{B}{0.25~\mathrm{G}}\right)^{-2}\left(\frac{\gamma}{70}\right)^{-1}\left(\frac{\delta}{1}\right)^{-1}~\mathrm{yr},
\end{equation}
where $\gamma$ is Lorentz factor of electron, $\delta$ is Doppler factor of moving blob. This time scale is a thousand times longer than $T_{flare}$ for both F1 and F2. 
Therefore, adiabatic expansion \citep{Laan(1966)} is considered to be one of the most dominant cooling mechanisms that can explain the cooling time scale of these outbursts.

    \begin{figure}[htb]
    \begin{center}
    \includegraphics[width=0.95\linewidth]{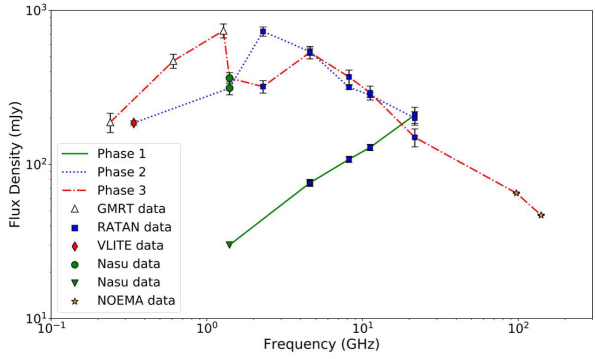}
    \end{center}
    \caption{
    Radio spectra of V404 Cygni in the quiescent (phase 1; solid line) and outburst phases (phase 2\&3: dotted and dashed lines, respectively) at 1.4\, GHz. Each color indicates radio spectra obtained in different phases. Filled green triangle shows an upper limit for 1.4\, GHz radio flux density in phase 1. Additionally, filled \textcolor{black}{red} diamond shows 0.34\, GHz radio flux density, but it was obtained more than 10 hours later compared to other radio flux densities in phase 2.}
    \label{fig:radio_spectra}
    \end{figure}
%
Figure \ref{fig:radio_spectra} compares radio spectra in the quiescent phase (phase 1) and outburst phases (phases 2 \& 3) by using archived data obtained within a couple of hours, including our 1.4\, GHz observation, as shown in Table \ref{tbl:tb2}. The time evolution of the radio spectra between the quiescent phase and the outbursts can be seen in this figure. Although the peak frequency ($\nu_\mathrm{p}$) in phase 1 is located higher than $\sim10$\,GHz, $\nu_\mathrm{p}$ in phases 2 \& 3 is located at around a few GHz. Such a change in spectral features possibly caused by non-thermal jet ejection is similar to that of the outburst detected in blazars at mm to sub-mm wavelengths (e.g., \cite{Tavares(2011)}), but the time scale was much shorter (i.e., within several days) compared to such extragalactic phenomena that occurred in the vicinity of supermassive black holes. 
Additionally, in phase 3, double peaks at $\sim1$\,GHz and $\sim5$\,GHz are clearly seen. On June 26, a flux density of $364\pm30$ mJy at 1.4 GHz was measured by the Nasu telescope.  As seen in the phase 3 line of figure 10, this Nasu data point at 1.4 GHz together with the nearby RATAN's one at 2.3 GHz  can be a strong support for the double-peak hypothesis.  As also discussed in \citet{Chandra(2017)}, there is a possibility that this double-peak spectrum implies a mixture of multiple synchrotron blobs ejected by outbursts.
Actually, a very long baseline interferometry (VLBI) follow-up observation carried out immediately after the X-ray outburst directly revealed the ejection of multiple non-relativistic jets and changing jet orientations \citep{Miller-Jones(2019)}. 
%
%
It is thought that the $\nu_\mathrm{p}$ of $\sim2$\,GHz in phase 2 shows the turn-over due to synchrotron-self absorption (SSA) of the non-thermal blob associated with the first 1.4\, GHz outburst seen in figure \ref{fig:long_light_curve}, which is probably related to the X-ray outburst detected on June 15. Then, it shifted to $\sim$\,1~GHz because of the change in the opacity of the ejected blob. The other $\nu_\mathrm{p}$ of $\sim4$\, GHz reflecting SSA turn-over due to a possibly newly ejected blob associated with the 2nd radio outburst occurred on June 26 (i.e., in phase 3).

\begin{table}[htbp]
\caption{Multi-frequency radio data in the quiescent and outburst phases.}\label{tbl:tb2}
\begin{center}
\scalebox{0.95}{
\begin{tabular}{cccccc}\hline\hline
Epoch & $\nu$ & $S$ & $S_\mathrm{err}$ & Date & Ref. \\
 & (GHz) & (mJy) & (mJy) & & \\\hline
Phase1 & 1.4 & $<30$ & - & 2015-Jun-18.73 & (1) \\
 & 4.6 & 76 & 4 & 2015-Jun-18.95 & (2) \\
 & 8.2 & 108 & 5 & 2015-Jun-18.95 & (2) \\
 & 11.2 & 129 & 6 & 2015-Jun-18.95 & (2) \\
 & 21.7 & 210 & 25 & 2015-Jun-18.95 & (2) \\\\
Phase2 & 0.34 & 186 & 6 & 2015-Jun-22.54 & (3) \\
 & 1.4 & 313 & 30 & 2015-Jun-21.73 & (1) \\
 & 2.3 & 730 & 50 & 2015-Jun-21.94 & (2) \\
 & 4.6 & 540 & 30 & 2015-Jun-21.94 & (2) \\
 & 8.2 & 317 & 10 & 2015-Jun-21.94 & (2) \\
 & 11.2 & 282 & 10 & 2015-Jun-21.94 & (2) \\
 & 21.7 & 200 & 20 & 2015-Jun-21.94 & (2) \\\\
Phase3 & 0.24 & 188 & 27 & 2015-Jun-26.89 & (4) \\
 & 0.61 & 470 & 49 & 2015-Jun-26.89 & (4) \\
 & 1.28 & 739 & 77 & 2015-Jun-26.89 & (4) \\
 & 1.4 & 364 & 30 & 2015-Jun-26.71 & (1) \\
 & 2.3 & 320 & 30 & 2015-Jun-26.93 & (2) \\
 & 4.6 & 534 & 50 & 2015-Jun-26.93 & (2) \\
 & 8.2 & 370 & 40 & 2015-Jun-26.93 & (2) \\
 & 11.2 & 292 & 30 & 2015-Jun-26.93 & (2) \\
 & 21.7 & 150 & 20 & 2015-Jun-26.93 & (2) \\
 & 97.5 & 65.2 & 0.2 & 2015-Jun-26.93 & (5) \\
 & 140.5 & 46.9 & 0.3 & 2015-Jun-26.93 & (5) \\\hline
\end{tabular}}
\end{center}

\begin{tabnote}
%
\textit{Column} 1: The state of source activity. \textit{Column} 2: Observed frequencies in GHz, \textit{Columns} 3\&4: Flux densities and errors, in mJy, \textit{Column} 5: Date on which each datum was obtained. Here we assume that radio data at low frequencies obtained within several hours show the same behavior. \textit{Column} 6: References. (1) Our observation, (2) \citet{Trushkin(2015)}, (3) \citet{Kassim(2015)}, (4) \citet{Chandra(2017)}, (5) \citet{Tetarenko(2015)}

\end{tabnote}
\end{table}

\section{Conclusion}
We used the Nasu telescope array, which is a spatial FFT interferometer, to carry out daily monitoring of radio transient phenomena by drift-scan observation. This telescope makes it possible to perform observation of all right ascension with a sensitivity of a few tens of mJy at a specific declination with a positional uncertainty of $\pm0.4$ deg over a day. On the other hand, no other radio observatory allows us to conduct wide-field daily monitoring with higher sensitivity especially in the East Asian region. Therefore, it is complementary to other radio telescopes, which can conduct not only monitoring of specific objects with high time resolution or high angular resolution, but also unbiased surveys with large FoV in other regions (e.g., \cite{chime(2018)}).

As a successful demonstration, our daily monitoring with the Nasu telescope from one month before the outburst of V404 Cygni occurred in June 2015 allowed us to capture the detailed behavior, which showed an abrupt change in flux density caused by an outburst in the 1.4\, GHz radio band.
The radio light curve showed two huge outbursts that occurred during a period of 10 days, and at least the 1.4\, GHz flux density during one of the two outbursts increased by more than ten times within a day, compared to the one in the quiescent state. We also have confirmed the extreme variation of radio spectra within the short period mentioned above by collecting other radio data observed at several radio observatories. Such spectral behaviors are considered to reflect the change in the opacity of the ejected blobs associated with radio and X-ray outbursts.

Although there is room to improve the calibration issues, which is crucial for accurate measurement of the amplitude and position of radio transients, as mentioned in \citet{Thompson(2017)}, this new type of wide-field radio telescope realized by adopting a spatial FFT technique is one of the most promising key techniques in the era of ``time-domain and multi-messenger astronomy''.

\begin{ack}
We are grateful to the anonymous referee, whose suggestions improved our paper substantially. This work was supported in part by Cooperative and Supporting Program for Research and Education in Universities of the National Astronomical Observatory of Japan (NAOJ) and by Japan Society for the Promotion of Science (JSPS) KAKENHI Grant Number JP15K05029 (K.T.), JP18H03721 (K.N.), JP15H00784 (K.N.). 
We thank Dr. Trushkin for providing us with their RATAN-600 data.

\end{ack}

\end{document}